## Research Article

# Lithography-Free, Low-Cost Method for Improving Photodiode Performance by Etching Silicon Nanocones as Antireflection Layer

**Jing Jiang, Zhida Xu, Jiahao Lin, and Gang Logan Liu**

*Department of Electrical and Computer Engineering, University of Illinois at Urbana-Champaign, Urbana, IL 61801, USA*

Correspondence should be addressed to Jing Jiang; jiang56@illinois.edu and Gang Logan Liu; loganliu@illinois.edu

A three-step process has been demonstrated to improve the performance of photodiode by creating nanocone forest on the surface of photodiode as an antireflection layer. This high-throughput, low-cost process has been shown to decrease the reflectivity by 66.1%, enhance the quantum efficiency by 27%, and increase the responsivity by 25.7%. This low-cost manufacture process can be applied to increase the responsivity of silicon based photonic devices.

## 1. Introduction

Photodetector is one of the most ubiquitous types of technology in use today. From a door opener to far-infrared cells on an astronomical satellite or the most advanced X-ray, photodetectors span a wide latitude of technologies. For the applications in everyday life, low-cost photodetectors have ample performance for needs like remote controls and automatic doors. For digital camera, though CMOS (complementary metal oxide semiconductor) is more popular than CCD (charge coupled device) due to its fast response, the low sensitivity to weak light limits its application at dark environment. Similarly, reflection also limits the performance for applications including infrared detectors, biological and chemical sensors, and solar cells [1–6]. Previously, researchers used antireflection coating which works well for a certain wavelength only [7, 8]. Using anisotropic etchants to form pyramids structures works only on crystalline silicon and is not suitable for nanosize devices. Moreover, at certain incident angles, the reflectivity is still high [9]. To improve the omnidirectional sensitivity within broadband wavelength, surface roughening and nanotexturing have been investigated [10–14]. Laser pulse irradiation in $SF_6$ environment [15] and self-mask reactive ion etching (RIE) [4, 16] can produce subwavelength structures on silicon and suppress the optical reflection below 5%. However, most of these methods are too expensive and complex for industrial application and their nanostructure sizes are as large as hundreds of nanometers. Since the emitter thickness of a photodiode is usually only hundreds of nanometers, traditionally, black silicon based photodiodes were fabricated starting from black silicon and formed PN junction on black silicon. If they directly apply the blackening process on a photodiode, the blackening processes will damage the PN junction.

In this paper, we demonstrated that a lithography-free, self-masked nanotexturization process produced tunable nanocone forest structure on a silicon based photodiode at room temperature. The performance has been improved after treatment.

## 2. Methods

Previously, our group has demonstrated that nanotexturization of silicon can improve the efficiency of solar cells, increase the sensitivity of IR sensing with bimaterial microcantilevers, and enhance the signal strength of surface-enhanced Raman spectroscopy (SERS) [3, 4, 16–18]. In this research, the black silicon was produced by a self-masked, lithography-free reactive ion etching (RIE) process. This three-step $O_2$-$CHF_3$-$Cl_2$ process at room temperature taking around 15 minutes can produce black silicon with different lengths reliably [3, 17]. Figure 1 shows the fabrication process and the scanning electron microscope (SEM) images of the nanotextured silicon. First, after degreasing process, wafer is placed into the chamber and a thin film of oxide on silicon is formed by oxygen plasma. This step takes 5 min (Figure 1(a)). At the second step, $CHF_3$ is flowed into the chamber and high power plasma is generated for 2 min. This short period of $CHF_3$ plasma etches the thin oxide layer to form dispersed silicon oxide islands instead of etching away the whole silicon dioxide layer (Figure 1(b)). At the third step, a mixture of $Cl_2$ and Ar with the ratio of 10 to 1 is turned on. This step is for etching the silicon and sculpturing the nanocones under the oxide nanomask formed at the second step (Figure 1(c)).

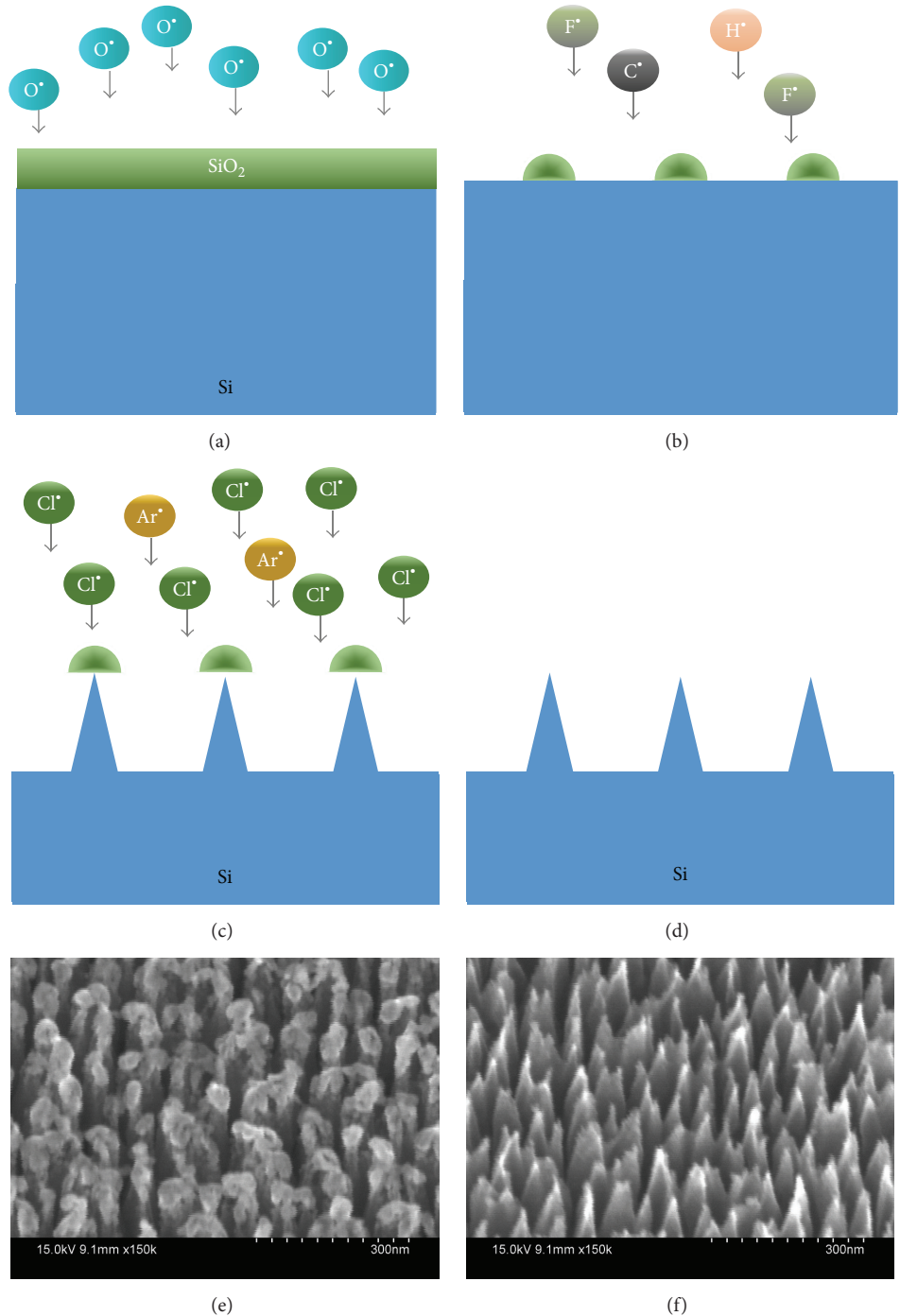

Figure 1: (a) A thin oxide layer (navy) formed on silicon surface (light blue) by oxygen plasma. (b) Dispersed oxide nanomask formed by etching thin oxide layer with $CHF_3$ plasma. (c) Nanocones etched by mixture plasma of $Cl_2$ and Ar (10 : 1). (d) Nanocone structure after silicon oxide removal by BOE. (e) SEM picture corresponding to (c). (f) SEM picture corresponding to (d).

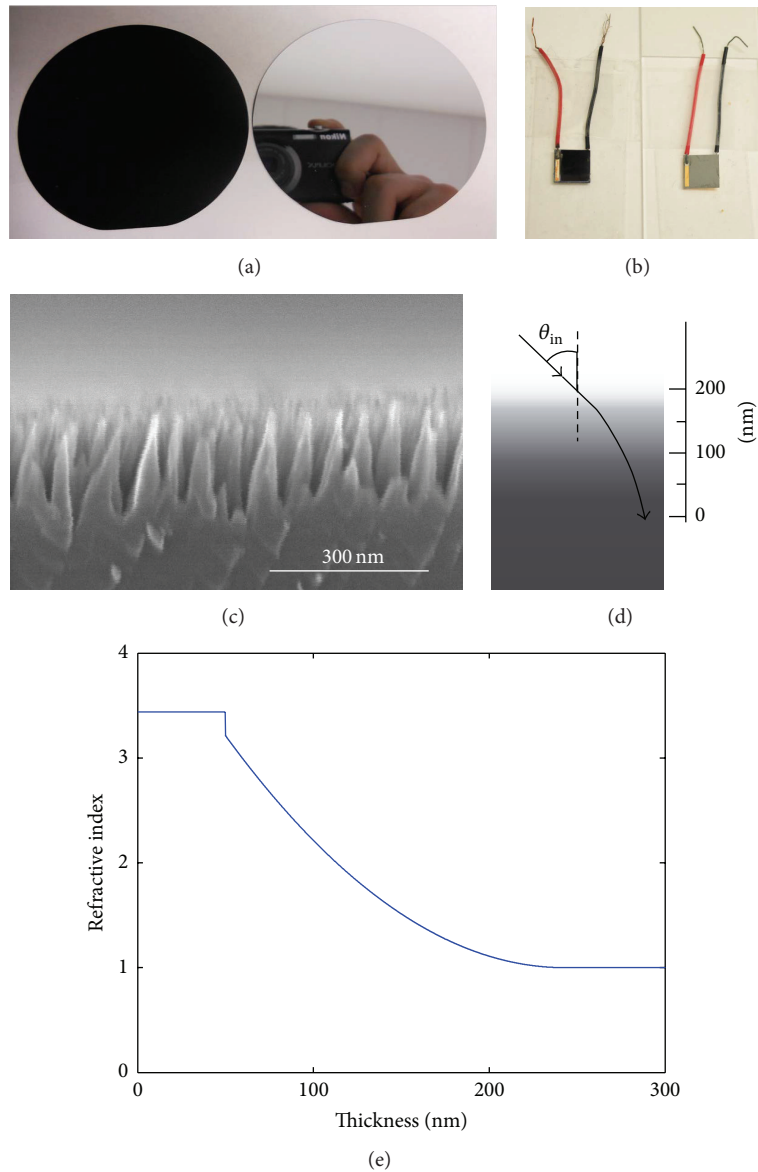

Figure 2: (a) Comparison of 3-inch blackened silicon (left) and polished silicon wafer (right). (b) Treated (left) and untreated (right) photodiode taped on a piece of glass slide. (c) Cross-sectional SEM image of blackened photodiode. (d) Effective refractive index of nanocone forest and the light path from air to silicon with nanocone structure, the reflection of light is minimal at each depth of the sample. (e) Schematic to show the effective refractive index of nanocone array on black silicon.

Chloride radicals generated in this step react with silicon chemically, and Argon can bombard silicon physically. The etching power and time are tunable at this step to control the length of nanocone structures. This three-step process can be finished in one chamber at room temperature. Figure 1(e) shows the SEM pictures of silicon sample after step 3. We can observe the oxide nanomasks and the nanocone structures beneath them. Finally, we can take the sample out of the chamber and dip it into buffered oxide etcher (BOE) for 5 seconds, the silicon nanomasks can be removed (Figure 1(d)), and Figure 1(f) shows the SEM image of nanocone structure after removing oxide nanomask.

After applying the above process on a piece of three-inch wafer, we can completely blacken it. As Figure 2(a) shows, the left one is a piece of three-inch silicon wafer after the three-step RIE process and the right one is original silicon wafer. We can see that, before the process, the polished silicon is very reflective while after the process, it turns totally black. Figure 2(b) shows the silicon based photodiode we used in this project. The right one was before treatment and the left one was after treatment. Each photodiode is taped on a piece of glass slide. We placed the photodiode into the chamber and after the three-step treatment process, the photodiode was successfully blackened as Figure 2(b), left side, shows.

Because the contact resistance is a very important parameter to minimize for PD manufacture, we protected that part by a layer of photoresist before performing the etching. From the picture (Figure 2(b)), we can observe that the gold contact electrode appears intact.

To explain the reason that blackening process can improve the responsivity of photodiode, let us take a look at the side view of blackened photodiode in Figure 2(c). We can observe that, on the surface of black photodiode, nanocones of around 190 nm height and 30 nm width are formed. According to the effective media theory, the gradually varying refractive index of the nanocone forest structure can dramatically reduce the reflection [19, 20]. According to Stephens and Cody's effective medium model, the composite materials have effective refractive index as averaging the values of the constituents that directly make up the composite material [21]. Thus, within the region of nanocones, the refractive index increases gradually and the light experienced minimum refection. At the interfaces of each depth, the equivalent impedances are very close to each other, so that impedances were matched at all the interfaces. If we observe the light path as shown in Figure 2(d), with incidence angle of $\theta_{in}$, the refractive angle decreases gradually as the light travels deeper into the nanocone structure with minimal reflection. The blue scale is the distance from solid silicon material. By assuming ideal close packed conic structure, we plot the estimated refractive index-thickness plot in Figure 2(e) by averaging the refractive index of the composite at each depth.

## 3. Results

The measurement of reflectivity, external quantum efficiency, and responsivity was performed with Cary 5000 UV-Vis-NIR spectrophotometer equipped with an integration sphere. External quantum efficiency (EQE) is the ratio of the number of charge carriers collected by the photodiode to the number of photons of a given energy shining on the photodiode from outside (incident photons):

$$\text{EQE} = \frac{\text{electrons/sec}}{\text{photons/sec}} \\ = \frac{\text{current/(charge of one electron)}}{(\text{total power of photons})/(\text{energy of one photon})}. \quad (1)$$

To convert from responsivity ($R_\lambda$, in A/W) to $\text{QE}_\lambda$ [22] (on a scale of 0 to 1),

$$\text{QE}_\lambda = \frac{R_\lambda}{\lambda}\frac{hc}{e} \approx \frac{1240 R_\lambda}{\lambda} \text{ (W} \cdot \text{nm/A)}, \quad (2)$$

where $\lambda$ is the wavelength in nm, $h$ is the Planck constant, $c$ is the speed of light in vacuum, and $e$ is the elementary charge.

Spectral responsivity measurements involve the measurement of the photocurrent produced by light of a given wavelength and power. The quantum efficiency is typically measured with bias light simulating reference conditions, because the device may be nonlinear. Typically, the spectral correction factor for efficiency measurements is calculated based on QE measurements near 0 V and is assumed to be the same as that at the maximum power point. This assumption is valid for most PV systems and results in a negligible error for amorphous silicon, which has a voltage-dependent spectral responsivity.

Figure 3(a) is the reflectance spectra on untreated and treated photodiodes. The sample is mounted in the center of the integrating sphere where the reflected and scattered light from the sample will be collected by the detector on the integrating sphere in order to measure reflectivity correctly. With the integrating sphere setup, the diffusive reflection from all angles was collected and sent to a spectrometer measuring wavelength ranging from 280 nm to 1100 nm. On average, the reflection of original photodiode is 5.3% while the reflection of black photodiode is only 1.8% which is 33.9% of the original photodiode reflection. We can see that the black photodiode overall suppresses the reflection across a broadband. When the wavelength is close to the band gap of silicon, the reflection started to increase when the wavelength is larger than 1,000 nm. To compare the quantum efficiencies of photodiode before and after the blackening treatment at each wavelength, we measured the external quantum efficiency (EQE) spectrum for the wavelength from 280 nm to 1100 nm plotted in Figure 3(b). Red curve is the EQE before treatment and the black curve is that after treatment. The average EQE for the photodiode before treatment was 52.4% and after treatment it was 66.5%. Overall, EQE was improved by 27.0% across the whole measured wavelength and was enhanced significantly between visible range and near-infrared range. This improvement is primarily due to the enhanced absorption by black silicon. In the end, the responsivity was measured in Figure 3(c). The black photodiode has 0.392 average responsivity while the untreated one was only 0.312. Thus, the blackening process generates 25.7% higher response at the whole measured wavelength range.

## 4. Discussion

*4.1. Selection of the Length of Nanocone.* According to Stephens and Cody's effective medium model, the reflectivity cutoff for textured silicon surfaces should be at a wavelength of four to six times larger than the depth of the textured layer [21]. Because the band gap of silicon is 1.11 eV and the cutoff wavelength is 1.13 μm, the length of nanocones should be around 188 nm to 325 nm. However, considering the fact that N-type top layer is only hundreds of nanometers thick, shorter nanocones are preferred so that the PN junction can be preserved. Thus, according to this design rule, our fabricated nanocones have length of 190 nm. In Figure 3(a), the reflectivity increased rapidly when the wavelength is larger than 1 μm and the reflectivity increases much faster with the increment of wavelength.

*4.2. Performance Improvement.* After calculation of the data shown in Figures 3(a)–3(c), we know that the average reflectance is suppressed by 66.1%, while the EQE and responsivity were only increased by 27.0% and 25.7%, respectively. From the data, we can tell that not all of the additional absorbed photons were transferred into electrical responses. There are two reasons associated with this phenomenon. First, the increased surface area introduces larger surface recombination rate; thus more generated electrons at surface were recombined due to the dangling bonds of silicon before diffusing to the electrode. To solve this issue, the surface can be treated under hydrogen environment so that the surface recombination effect can be reduced. Second, as more photons are absorbed, the portion of energy that cannot transfer to electricity will change into the form of phonons, eventually becoming heat dissipation. Consequently, the temperature becomes higher during the illumination. At higher temperature, the band gap became smaller and photons with long wavelength cannot generate electrons. If we observe Figures 3(b) and 3(c) closely, the treated photodiode has lower QE and responsivity around 1100 nm wavelength. However, with some imperfection, with this simple treatment method, we can increase the responsivity by 25.7%. As a result, it can measure photons at lower intensity.

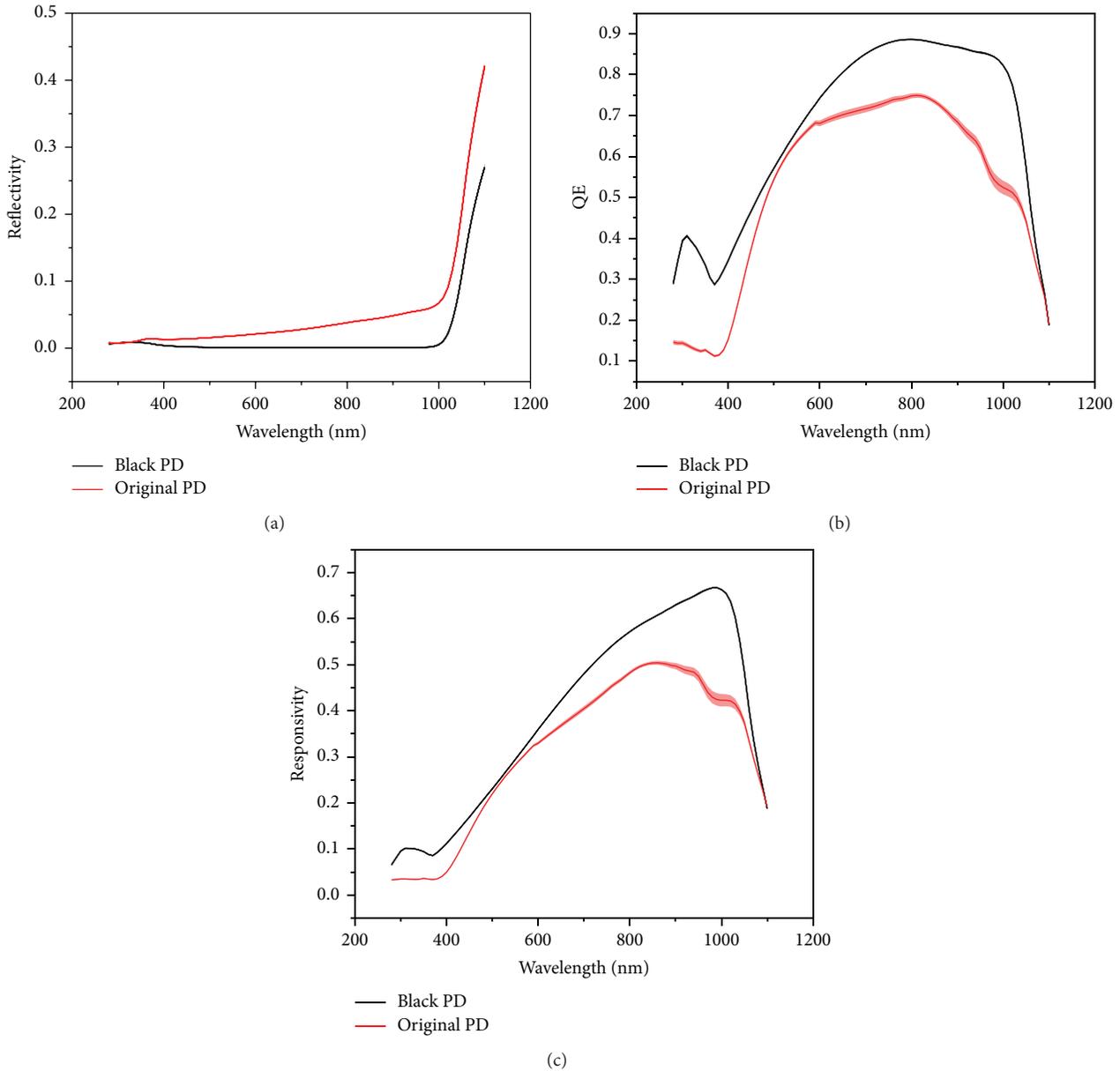

Figure 3: (a) Diffusive reflection spectra of untreated PD (red) and treated black PD (black). (b) External quantum efficiency of untreated PD (red) and treated black PD (black). (c) Responsivity of untreated PD (red) and treated black PD (black). Shaded area is standard deviation.

*4.3. Potential Application.* The three-step blackening treatment in this research can be applied to IR camera, low optical signal sensing, and so on. This process can tune the length of nanocone structure easily in order to optimize the design of photonic sensors targeting for different wavelength. Within 15 minutes of treatment, the performance of optical detector can increase by 25.7% by utilizing a single piece of equipment. Compared with other techniques, this method is cost-effective. In addition, this method is quite expandable for high-throughput manufacture. During our research, this method has already shown more than 97% yield during research stage. It is highly tunable and is capable of producing very consistent results for planar surface, single-crystalline, and multi-crystalline silicon with low requirement to the manufacture condition (Supplementary Material available online at http://dx.doi.org/10.1155/2016/4019864). Thus, it is highly applicable for industrial application. At the cost side, the fixed cost is the equipment for inductive coupled plasma reactive ion etching, and the variable cost is the gas and electricity. It can either be integrated with current manufacturing pipeline with this additional equipment or utilize current equipment to fulfill the blackening recipe to achieve the improvement. In addition, this process can be applied after forming metal contact. Thus, it can be used for both creating new devices and improving old devices.

## 5. Conclusion

We have demonstrated that, after being nanotexturized by the three-step blackening process for about 15 minutes, the photodiode became more sensitive. The reflectivity was suppressed by 66.1%, the EQE was increased by 27.0%, and the responsivity was improved by 25.7%. This low-cost, high-throughput, rapid nanotexturization method has the potential to improve the efficiency of silicon photodiode devices in semiconductor industry.

## Competing Interests

The authors declare that they have no competing interests.

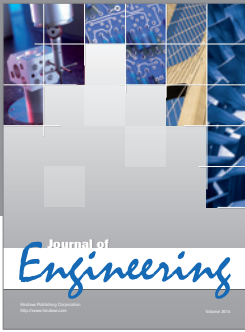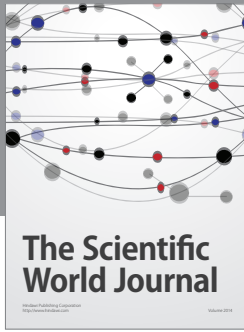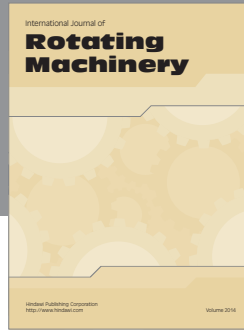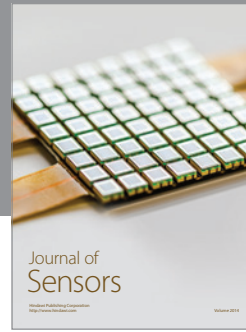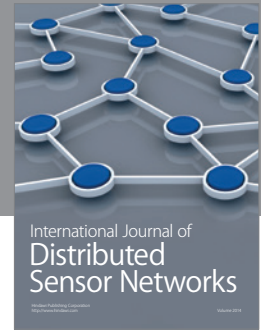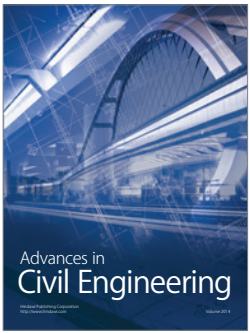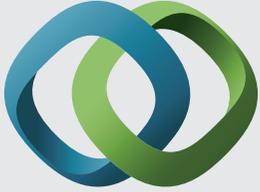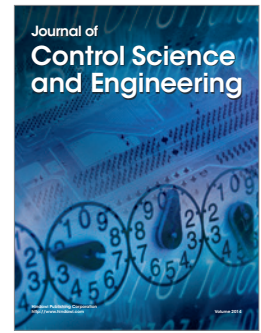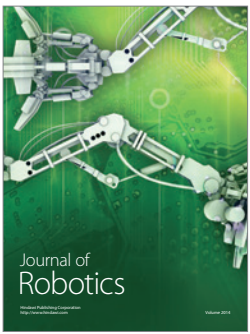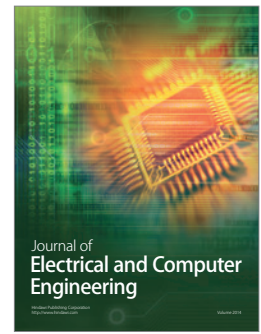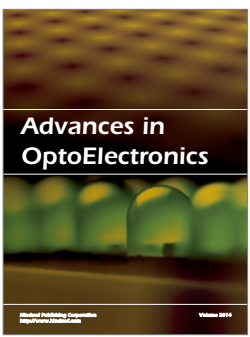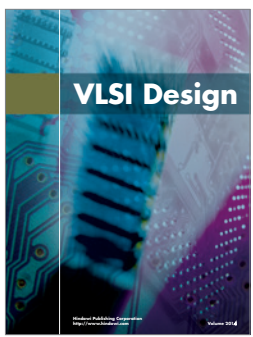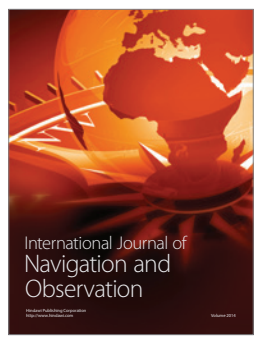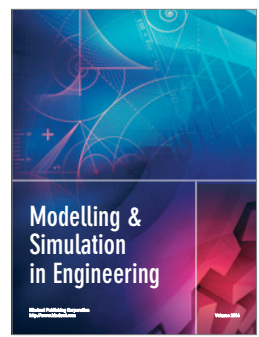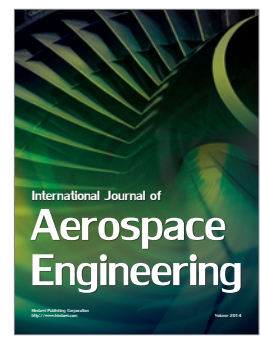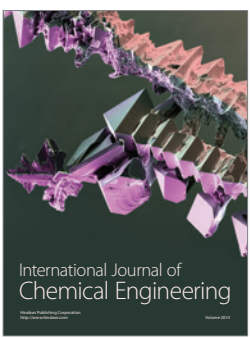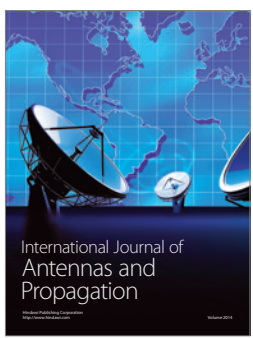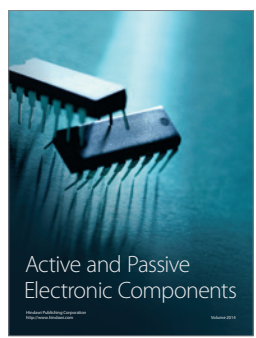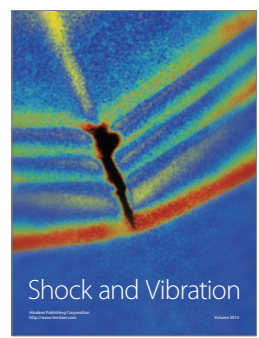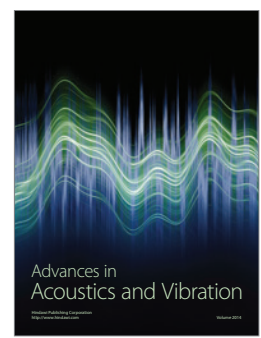